\documentclass[aps,pra,twocolumn,superscriptaddress,longbibliography]{revtex4-2}

\usepackage{graphicx}
\usepackage{amsmath,amssymb}
\usepackage{xcolor}
\usepackage{hyperref}
\hypersetup{colorlinks=true, linkcolor=blue, citecolor=blue, urlcolor=blue}

\newcommand{\rihead}[1]{\smallskip\noindent\textit{#1.---}}

\begin{document}

\title{Machine learning classification of dynamical phases in a driven nonlinear\\
Su-Schrieffer-Heeger lattice}

\author{Chukwunaedum Patrick Izediuno}
\affiliation{School of Physical, Chemical and Environmental Sciences, Cardiff University, Cardiff CF24 3AA, United Kingdom}
\author{Ghada H. Alharbi}
\affiliation{Department of Physics, University of Tabuk, Tabuk 741, Kingdom of Saudi Arabia}
\author{Stephan Wong}
\affiliation{Ecole Centrale de Lyon, CNRS, INSA Lyon, Universit\'e Claude Bernard Lyon 1,
CPE Lyon, CNRS, INL, UMR5270, 69130 Ecully, France}
\author{Zeeshan Ahmad}
\affiliation{Bentham Instruments Ltd, Reading RG2 0NH, United Kingdom}
\author{Sunae So}
\affiliation{Department of Control and Instrumentation Engineering, Korea University, 2511, Sejong-ro, Sejong, 30019 Republic of Korea}
\affiliation{Department of Electro-Mechanical System Engineering, Korea University, 2511, Sejong-ro, Sejong, 30019 Republic of Korea}
\author{Minkyung Kim}
\affiliation{Department of Mechanical and Robotics Engineering, Gwangju Institute of Science and Technology (GIST), 61005 Gwangju, Republic of Korea}
\author{Sang Soon Oh}
\email{ohs2@cardiff.ac.uk}
\affiliation{School of Physical, Chemical and Environmental Sciences, Cardiff University, Cardiff CF24 3AA, United Kingdom}

\date{\today}

\begin{abstract}
Driven nonlinear photonic lattices can settle into steady, symmetry-broken, periodic, or chaotic states, and mapping where each occurs in parameter space is key to device design. Brute-force mapping is slow and relies on hand-built criteria, while machine-learning classifiers are usually tested on different systems, so it is unclear which suits a new lattice. 
We therefore apply four classifiers to the same coherently driven Su-Schrieffer-Heeger (SSH) chain of Kerr resonators: a hand-built heuristic classifier based on Lyapunov exponents and spectral diagnostics, a supervised convolutional neural network (CNN) trained on the heuristic labels, an unsupervised library-building scheme based on augmented dynamic mode decomposition (aDMD), and an unsupervised Gaussian mixture model (GMM) on four dynamical invariants that include an aDMD reconstruction error. Using a converged heuristic diagram as the reference, the CNN reaches $98.0\%$ held-out accuracy, while the two label-free methods reach adjusted Rand indices (ARIs) of 0.916 and 0.947, respectively, without seeing any labels. 
We show that the aDMD library cleanly resolves the stable and bistable phases and separates symmetric from symmetry-broken periodic oscillations, which the heuristic misses, though it cannot distinguish periodic from chaotic switching. 
We also find that a poorly converged reference understates the label-free methods' accuracy. 
\end{abstract}

\maketitle

\rihead{Introduction}
Arrays of coupled nonlinear resonators support a wide range of dynamical behaviour, including stable steady states, spontaneous symmetry breaking, self-sustained oscillations, and chaos~\cite{strogatz2015,ikeda1979,longhi2018,woodley2021}. Such arrays are realised in topological insulator lasers~\cite{bandres2018,harari2018}, coupled ring-resonator and microresonator systems~\cite{tikan2021,alharbi2024}, and polariton condensates~\cite{ohadi2017}. Even a single Kerr ring resonator that is driven equally in both directions shows this behaviour: above a threshold drive, the mirror symmetry between its clockwise and counter-clockwise fields is spontaneously broken, so that one circulation direction carries more light than the other, and at higher drive the dominant direction can alternate periodically in time~\cite{delbino2017,woodley2021}. For all of these platforms, a basic task is to construct the phase diagram, that is, to determine which type of long-time behaviour occurs at each point of the driving-parameter space. Analytical approaches are restricted to simple or weakly nonlinear cases~\cite{longhi2018,malzard2018njp}. Numerical sweeps are more general, but they become expensive when neither the number of phases nor the location of their boundaries is known in advance, because every candidate behaviour must be anticipated and tested for explicitly.

Machine learning (ML) offers an alternative route to phase classification in physics~\cite{carleo2019}. Supervised neural networks have identified phases in spin models~\cite{carrasquilla2017} and topological band structures~\cite{zhang2018}, and unsupervised methods such as autoencoders~\cite{wetzel2017}, confusion schemes~\cite{vannieuwenburg2017}, and clustering of topological features~\cite{rodrigueznieva2019,scheurer2020} have located transitions without labelled data. For driven systems, where each parameter point is described by a time series rather than a static configuration, dynamic mode decomposition (DMD)~\cite{schmid2010,tu2014} is a natural tool, because it fits the data with a small set of modes that each oscillate and grow or decay at a fixed rate. Its time-augmented variant (aDMD)~\cite{kramer2017} was used by Wong~\textit{et al.}~\cite{wong2023,wong2026} as the basis of a label-free method for drawing the phase diagram of a topological SSH laser with saturable gain. Convolutional neural networks (CNNs) have been used to classify phases in nonlinear polariton lattices~\cite{zvyagintseva2022}, and neural networks have also been applied to inverse design in topological photonics~\cite{pilozzi2018}.

The available methods differ mainly in what they require from the user. A heuristic classifier, by which we mean a fixed set of hand-chosen diagnostics and thresholds, must be told in advance what every phase looks like. A supervised CNN needs no such description, but it needs labelled training data, and for a new system these labels usually come from a heuristic classifier run on part of the parameter space. Label-free methods need neither, but they return clusters whose physical meaning must be established afterwards, and whose number need not match the number of phases one would typically draw by hand. Because each method has so far been demonstrated on a different system, with different simulation settings and figures of merit, these trade-offs have not been quantified side by side. There is therefore little guidance on the choice of method for treating a new nonlinear lattice, or on what each method will and will not reveal.

In this work we provide that comparison for a driven SSH lattice of Kerr-nonlinear ring resonators, a system whose edge modes are known to show bistability and periodic and chaotic switching~\cite{alharbi2024}. We apply four classifiers to the same simulated data (Table~\ref{tab:methods}): (i) a heuristic classifier, which serves as the reference; (ii) a supervised CNN trained on 70\% of the heuristic labels; (iii) an aDMD library-building scheme that follows Refs.~\cite{wong2023,wong2026}; and (iv) a Gaussian mixture model (GMM) on four dynamical invariants that include an aDMD reconstruction error. Methods (ii)--(iv) span the range from fully supervised to fully label-free, so that the comparison shows what is gained and lost as labels are removed. We compare the methods in terms of agreement with the reference, labelling requirement, computational cost, and interpretability, and we check whether the extra structure found by the label-free methods reflects physics or artefacts. We find that both DMD-based methods recover the main phase boundaries without labels; that the library method separates symmetric from symmetry-broken oscillations, which the heuristic groups together, but merges periodic and chaotic switching; and that a short simulation window can bias the reference itself, so that convergence of the reference must be checked before any classifier is scored against it.

\begin{figure}[t]
\centering
\includegraphics[width=\linewidth]{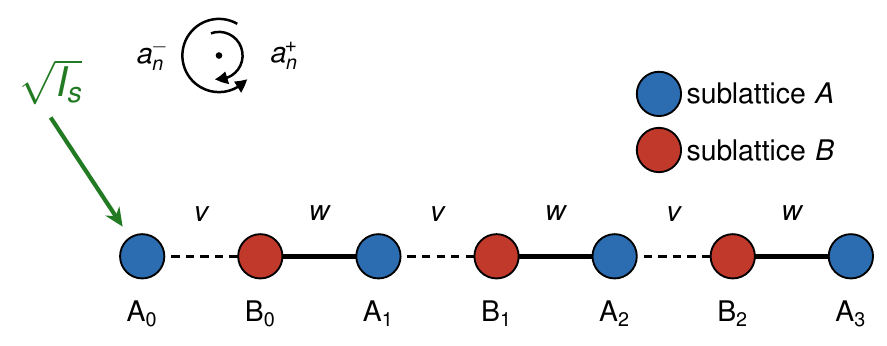}
\caption{Schematic of the driven SSH chain with $M=7$ ring resonators. Sublattice-$A$ (blue) and sublattice-$B$ (red) sites are coupled by intra-cell hopping $v$ (dashed) and inter-cell hopping $w>v$ (solid). The first site is driven coherently with amplitude $\sqrt{I_s}$, and every site carries two counter-propagating field components, $a_n^{\pm}$ or $b_n^{\pm}$.}
\label{fig:schematic}
\end{figure}

\rihead{Model}
We consider a one-dimensional SSH chain~\cite{su1979} of $M=7$ Kerr-nonlinear ring resonators (Fig.~\ref{fig:schematic}), with $N+1=4$ sites on sublattice $A$ and $N=3$ on sublattice $B$, intra-cell coupling $v$, and inter-cell coupling $w$. For $w>v$, used throughout, the chain is in its topological phase and the unpaired boundary site supports a localised edge mode~\cite{malzard2018oe,zhao2018}. Each ring carries a clockwise and a counter-clockwise field, $a_n^{\pm}$ on sublattice $A$ and $b_n^{\pm}$ on sublattice $B$, which are coupled through self- and cross-phase modulation. In an isolated ring driven equally in both directions, this coupling makes the symmetric state unstable above a threshold drive, so that one direction becomes dominant~\cite{delbino2017,hill2020}; at higher drive the dominant direction can alternate periodically in time~\cite{woodley2021}. Refs.~\cite{delbino2017,hill2020} give an introduction to this single-resonator model. In the lattice, both field components of the first $A$ site are driven coherently with equal amplitude $\sqrt{I_s}$, where the drive intensity $I_s$ absorbs the input-coupling efficiency. Following Ref.~\cite{alharbi2024}, the equations of motion are
\begin{align}
\dot a_n^{\pm} &= -a_n^{\pm} - i\Delta\, a_n^{\pm} + i\!\left(\alpha|a_n^{\pm}|^2 + \beta|a_n^{\mp}|^2\right) a_n^{\pm} \notag\\
&\quad + i v\, b_n^{\mp} + i w\, b_{n-1}^{\mp} + \sqrt{I_s}\,\delta_{n,0}, \label{eq:eom_a}\\
\dot b_n^{\pm} &= -b_n^{\pm} - i\Delta\, b_n^{\pm} + i\!\left(\alpha|b_n^{\pm}|^2 + \beta|b_n^{\mp}|^2\right) b_n^{\pm} \notag\\
&\quad + i v\, a_n^{\mp} + i w\, a_{n+1}^{\mp}, \label{eq:eom_b}
\end{align}
where $\Delta$ is the detuning and $\alpha$ and $\beta$ are the self- and cross-phase modulation coefficients. All rates are expressed in units of the resonator decay rate, so time is measured in units of the inverse decay rate. We fix $\alpha=1$, $\beta=4$, $v=3$, and $w=7$, and sample $(\Delta,I_s)\in[-10,10]\times[0,50]$ on a uniform $60\times60$ grid. Each trajectory starts from a near-symmetric state with a $1\%$ imbalance between $a_0^{+}$ and $a_0^{-}$, and is integrated with a fourth-order Runge-Kutta scheme (step $\delta t=0.02$) up to $t_f=120$. The first half is discarded as a transient and the remainder is downsampled by a factor of three. We chose this window after checking convergence: rerunning the heuristic classifier with $t_f=240$ changes only $1.6\%$ of the labels, whereas the shorter window $t_f=60$ with the first third discarded changes $7.3\%$, mostly slowly decaying oscillations that are later classified as bistable (Appendix~\ref{app:conv}). All methods use the same window.

\begin{figure}[t]
\centering
\includegraphics[width=\linewidth]{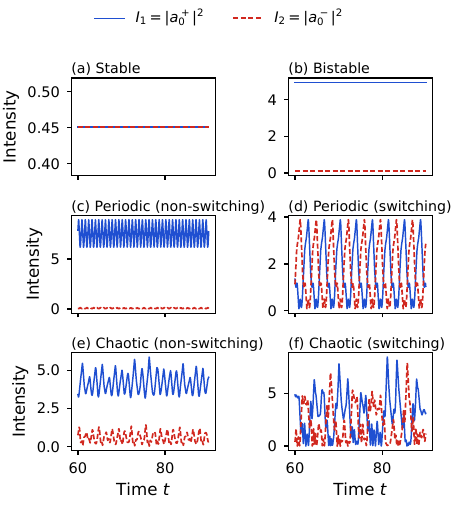}
\caption{Representative long-time dynamics of the two field intensities at the driven site, $I_1=|a_0^{+}|^2$ (solid) and $I_2=|a_0^{-}|^2$ (dashed), for each of the six phases, taken from the interior of each region of the reference diagram. (a) Stable, $(\Delta,I_s)=(-3.22,21.19)$: both intensities are constant and equal, so the two curves coincide. (b) Bistable, $(2.88,28.81)$: constant but unequal intensities. (c) Periodic, non-switching, $(5.93,43.22)$: periodic oscillation with $I_1>I_2$ throughout. (d) Periodic, switching, $(6.27,16.10)$: the dominant component alternates. (e) Chaotic, non-switching, $(7.97,40.68)$. (f) Chaotic, switching, $(8.64,31.36)$.}
\label{fig:dynamics}
\end{figure}

Depending on $(\Delta,I_s)$, the lattice settles into one of four types of long-time behaviour, shown at the driven site in Fig.~\ref{fig:dynamics}. In the stable phase [Fig.~\ref{fig:dynamics}(a)] both intensities approach the same constant value, as expected from the symmetric drive. In the bistable phase [Fig.~\ref{fig:dynamics}(b)] the intensities are again constant but unequal, $I_1\neq I_2$: the mirror symmetry between the two circulation directions is spontaneously broken, as in a single ring~\cite{delbino2017,hill2020}, and the mirror state with $I_1$ and $I_2$ exchanged is equally stable, which is why we call this phase bistable. In the periodic phases [Fig.~\ref{fig:dynamics}(c,d)] the intensities oscillate with a well-defined period, and in the chaotic phases [Fig.~\ref{fig:dynamics}(e,f)] they fluctuate aperiodically. The oscillatory phases are further divided according to which field component is dominant, i.e.\ has the larger instantaneous intensity at the driven site. If the same component stays dominant throughout the sampled window [panels (c,e)], we call the state non-switching; if dominance passes back and forth between $a_0^{+}$ and $a_0^{-}$ [panels (d,f)], as in the self-switching oscillations of a single ring~\cite{woodley2021}, we call it switching. This gives six phases in total, which are the classes used by the heuristic and supervised classifiers below.

\rihead{Methods}
The four classifiers operate on the same post-transient trajectories, one for each of the 3600 grid points. The heuristic and aDMD + GMM each use one additional run per point, from a perturbed and a mirrored initial condition, respectively, and the baseline uses three runs from rescaled initial conditions. Table~\ref{tab:methods} lists their inputs, labelling requirements, and outputs, and Appendices~\ref{app:heur}--\ref{app:ae} give all thresholds and hyperparameters. We describe the methods in order of decreasing prior knowledge about the phases, from the hand-built heuristic to a baseline that uses neither labels nor physics-based features.

\begin{table}[t]
\caption{Overview of the classification methods compared in this work; full inputs are described in the text. All methods use the same simulation settings. LLE, largest Lyapunov exponent; CNN, convolutional neural network; aDMD, time-augmented dynamic mode decomposition; GMM, Gaussian mixture model; DP, Dirichlet process; ConvAE, convolutional autoencoder; IDEC, improved deep embedded clustering.}
\label{tab:methods}
\begin{ruledtabular}
\footnotesize
\begin{tabular}{llll}
Method & Input & Labels & Output \\
\hline
(i) Heuristic & LLE, peaks, switching & none & 6 phases \\
(ii) CNN & time series, spectra & $70\%$ of (i) & 6 phases \\
(iii) aDMD library & aDMD bases & none & regimes \\
(iv) aDMD + GMM & 4 invariants & none & $K$ (DP) \\
\hline
ConvAE + IDEC & time series & none & 6 clusters \\
\end{tabular}
\end{ruledtabular}
\end{table}

\textit{(i) Heuristic classifier.}---Each trajectory is assigned to one of the six phases using four diagnostics computed from the sampled window: the largest Lyapunov exponent (LLE), estimated with the two-trajectory method of Benettin~\textit{et al.}~\cite{benettin1980}; the standard deviation of the two intensities at the driven site; the number of prominent peaks in these intensities; and the number of times dominance passes between $a_0^{+}$ and $a_0^{-}$. An LLE above a small positive threshold identifies chaos. Otherwise, a trajectory is periodic if its intensities are not constant and show at least two prominent peaks, and a constant trajectory is bistable if its two mean intensities differ by more than $5\%$ and stable if they do not. A periodic or chaotic trajectory is labelled switching if the sign of $I_1-I_2$, counted only when $|I_1-I_2|$ exceeds $10\%$ of the mean intensity, changes at least twice in the sampled window. The result is used as the reference for all other methods. It is the most demanding method in terms of prior knowledge, since each phase it can report must be anticipated and given an explicit rule.

\textit{(ii) Supervised CNN.}---A convolutional network with three input branches is trained on a uniformly random $70\%$ of the grid points (drawn with a fixed random seed, without stratification by phase) using the heuristic labels, and is then applied to the remaining $30\%$, which are never seen during training. The first branch applies one-dimensional convolutions to the intensity time series of all 14 field components, and the second branch processes their log-power spectra; between them, these branches separate oscillatory from constant behaviour and periodic from chaotic spectra. The third branch receives the two mean intensities at the driven site and their normalised difference. It is needed because the stable and bistable phases are both constant in time and are distinguished only by whether the two mean intensities differ, information that is removed when each time series is normalised for the first branch. The branches are concatenated and passed to a fully connected classifier. Training uses the AdamW optimiser~\cite{loshchilov2019} with a cosine-annealed learning rate, a class-weighted cross-entropy loss to compensate for the very different sizes of the six phases, and label smoothing~\cite{goodfellow2016}, which discourages overconfident predictions near phase boundaries.

\textit{(iii) aDMD library building.}---DMD approximates a sequence of snapshots by a linear map, whose eigenvectors (the DMD eigenmodes) and eigenvalues give the spatial patterns, frequencies, and growth or decay rates present in the data~\cite{schmid2010,tu2014}. In aDMD, each DMD mode $\psi$ is stacked with copies of itself advanced in time by its eigenvalue $\lambda$, $(\psi,\lambda\psi,\lambda^2\psi,\dots)$, so that the resulting basis describes short stretches of the trajectory rather than single instants~\cite{kramer2017}. Following Refs.~\cite{wong2023,wong2026}, we build an aDMD basis from the raw complex field amplitudes of a trajectory and use it to decide whether other trajectories show the same dynamics, by checking how well the basis reconstructs them. In a bottom-up pass, trajectories are visited in random order, and a new basis is added to the library whenever none of the existing bases reconstructs the current trajectory to within a relative error $\epsilon_{\rm th}=0.02$. In a top-down pass, bases whose subspaces overlap by more than a threshold $\gamma_{\rm th}$ are merged into a single regime, and every trajectory is then assigned to the regime that reconstructs it best. The resulting library therefore contains as many regimes as are needed to describe all trajectories to the chosen tolerance, rather than a number fixed in advance. Because the bottom-up pass depends on the visiting order, we repeat it for three random orders and combine the three partitions by evidence accumulation~\cite{fred2005}: two grid points are placed in the same final regime if they share a regime in at least two of the three runs.

\textit{(iv) aDMD + GMM.}---The second label-free method summarises each trajectory by four scalar invariants and clusters them. The first is the aDMD reconstruction error: an aDMD basis is built from the trajectory exactly as in method (iii), and its relative error in reconstructing short windows of the same trajectory is recorded. This error is small for constant and periodic dynamics, which a low-rank basis describes well, and large for chaos. The second and third are the energy-weighted mean oscillation frequency and the energy-weighted mean growth or decay rate of the eigenvalues of the DMD from which the basis is built~\cite{schmid2010}. The fourth is a symmetry probe, obtained by repeating the simulation from the mirrored initial condition (with the imbalance between $a_0^{+}$ and $a_0^{-}$ reversed) and measuring the distance between the two sets of time-averaged intensities, which is non-zero only when the symmetry is broken. Before clustering, the reconstruction error is floored at $10^{-6}$, because smaller values are set by numerical round-off rather than by the dynamics (Appendix~\ref{app:gmm}). The standardised invariants are clustered with a Dirichlet-process GMM~\cite{mclachlan2000,blei2006} with up to 30 components. Its prior on the mixture weights favours few components, so components that the data do not support are left empty and the number of clusters is inferred rather than fixed in advance. For comparison, we also fit finite GMMs with $K=2$ to $30$ components and select $K$ by the Bayesian information criterion (BIC)~\cite{schwarz1978}, a standard statistical model-selection score (not to be confused with bound states in the continuum).

\textit{Baseline.}---To test whether hand-chosen DMD features are needed at all, we also cluster the latent representation of a one-dimensional convolutional autoencoder (ConvAE). The autoencoder is trained, without labels, to reproduce the intensity time series of all 14 field components through a 32-dimensional bottleneck, and its latent space is then refined with the improved deep embedded clustering (IDEC) loss~\cite{guo2017}, which adds a clustering term to the reconstruction loss. Each grid point is simulated from three slightly different initial conditions and assigned the majority cluster of its three time series. This method uses no physics-based features, but unlike the DMD-based methods it is given the number of clusters, which we set to the six reference phases.

\textit{Comparison metric.}---Methods (iii) and (iv) and the baseline return cluster labels whose number and numbering are arbitrary, so they cannot be scored by accuracy. We instead use the adjusted Rand index (ARI)~\cite{hubert1985}. For every pair of grid points, it checks whether the two labellings agree, either by placing both points in the same group or by placing them in different groups. The fraction of agreeing pairs is then corrected for chance: it is rescaled so that its expected value is 0 when one labelling is replaced by a random assignment of the points to groups of the same sizes, and its value is 1 when the two labellings are identical. The ARI does not require the two labellings to have the same number of clusters, which allows a partition into many regimes to be scored against the six reference phases. Its scale is not linear in the fraction of misassigned points, so small differences in ARI should not be over-interpreted. Because a high ARI can arise either from a near-identical partition or from a finer subdivision of the reference phases, we also report homogeneity, which is 1 when every cluster contains points from only one reference phase, and completeness, which is 1 when every reference phase lies within a single cluster.

\begin{figure*}[t]
\centering
\includegraphics[width=\textwidth]{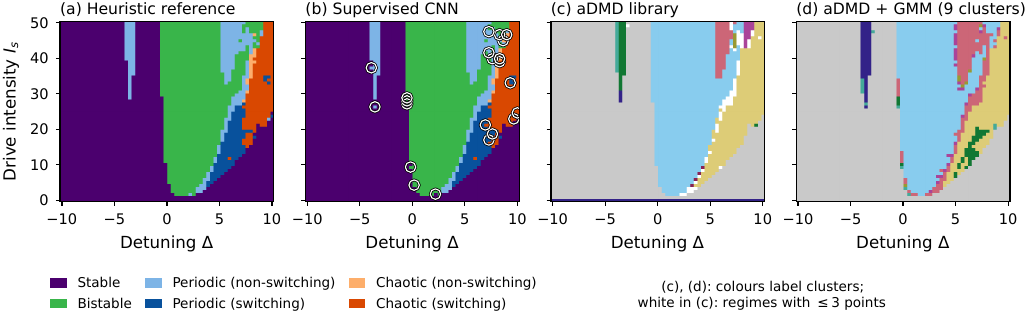}
\caption{Phase diagrams of the driven SSH lattice from the four methods. (a) Heuristic reference ($t_f=120$). (b) Supervised CNN: training points carry their reference labels and the held-out $30\%$ carry the network's predictions; open rings mark the 22 misclassified held-out points, with the predicted colour visible inside each ring. (c) aDMD library, consensus of three discovery orders; colours label the ten regimes with more than three points, and white marks the remaining 50 small regimes. (d) aDMD + GMM, Dirichlet-process mixture with 9 clusters. In (c) and (d) the colours identify clusters only and do not correspond to the phase colours of (a) and (b).}
\label{fig:diagrams}
\end{figure*}

\rihead{Results}
Figure~\ref{fig:diagrams} shows the phase diagrams obtained by the four methods, and Table~\ref{tab:comparison} summarises their agreement with the reference. The reference diagram [Fig.~\ref{fig:diagrams}(a)] is dominated by the stable phase at negative detuning and at weak drive. Bistability occupies a wedge that opens at positive detuning above a threshold drive of $I_s\approx2$, consistent with symmetry breaking of the edge-site fields reported in Ref.~\cite{alharbi2024}. At larger detuning the bistable wedge gives way to periodic states, with the switching periodic phase on its lower edge, and chaotic switching dynamics appear at the largest detuning and drive, in line with the instabilities known for driven Kerr cavities~\cite{ikeda1979}. The chaotic non-switching phase occupies only 17 grid points, scattered along the boundary between the periodic and chaotic regions. A narrow strip of periodic non-switching states also appears inside the stable region near $\Delta\approx-3.5$ for $I_s\gtrsim28$.

\begin{table}[b]
\caption{Agreement of each method with the heuristic reference on the full $60\times60$ grid. For the CNN, accuracy is measured on the 1080 held-out points; for the label-free methods, ARI, homogeneity (Hom.), and completeness (Comp.) are computed over all 3600 points.}
\label{tab:comparison}
\begin{ruledtabular}
\begin{tabular}{lcccc}
Method & Agreement & Hom. & Comp. & Clusters \\
\hline
(ii) CNN & $97.96\%$ (acc.) & --- & --- & 6 \\
(iii) aDMD library & $0.916$ (ARI) & $0.865$ & $0.760$ & $60^{a}$ \\
(iv) aDMD + GMM & $0.947$ (ARI) & $0.899$ & $0.839$ & 9 \\
ConvAE + IDEC & $0.375$ (ARI) & $0.679$ & $0.470$ & $6^{b}$ \\
\end{tabular}
\end{ruledtabular}
\footnotesize
$^{a}$ 10 regimes with more than three points. \\
$^{b}$ Number of clusters fixed in advance.
\end{table}

Trained on $70\%$ of the reference labels, the CNN reproduces the reference diagram closely [Fig.~\ref{fig:diagrams}(b)], with a held-out accuracy of $97.96\%$ (22 of 1080 points misclassified). This level of agreement is expected, since the network learns from the reference itself. Twenty-one of the 22 errors lie on grid points next to a phase boundary of the reference, although such points make up only $17\%$ of the grid, so the errors reflect uncertainty in where a boundary lies rather than confusion between unrelated phases. The network also separates the stable and bistable phases, which differ only in their mean intensities, showing that the scalar input branch carries this information as intended. The choice of $70\%$ for the training set is not critical. With the held-out points kept fixed and five random training subsets of each size, the mean held-out accuracy is $98.3\%$ for training sets of $50\%$ or $70\%$ of the grid, $97.9\%$ for $30\%$, and still $95.5\%$ when only $5\%$ of the grid (180 points) is labelled, with most errors again on phase boundaries (Appendix~\ref{app:cnn} and Fig.~\ref{fig:frac}). This overall accuracy is dominated by the large stable and bistable phases, however, and the small oscillatory phases are recognised less reliably when fewer labels are available (Appendix~\ref{app:cnn}). The CNN is therefore an efficient way to extend a known classification to new parameter points, but it cannot report a phase that is absent from its training labels.

The aDMD library [Fig.~\ref{fig:diagrams}(c)] needs neither labels nor a prescribed number of phases. Individual discovery orders give between 24 and 187 regimes, because the automatically selected merging threshold depends on the order, but their partitions agree well with one another (pairwise ARI $0.91$--$0.95$). The consensus of the three orders contains 60 regimes and reaches ${\rm ARI}=0.916$ against the reference, with homogeneity $0.865$ and completeness $0.760$. Only ten of these regimes contain more than three grid points. The remaining fifty hold 67 points ($1.9\%$ of the grid), 55 of which lie next to a phase boundary of the reference. A single discovery run with the strictest threshold fragments the oscillatory phases far more, giving many periodic and chaotic trajectories a regime of their own. Within a periodic phase, these fragments correspond to narrow bands of oscillation frequency (Appendix~\ref{app:admd}). They depend on the discovery order and largely disappear in the consensus, so they should not be read as additional phases.

The ten substantive regimes relate to the reference as follows. The stable and bistable phases each fall almost entirely into one regime ($96\%$ and $98\%$ of their points), and the undriven row $I_s=0$, where all fields decay to zero, forms a regime of its own. All switching dynamics, periodic and chaotic, are grouped into a single regime, so the library does not distinguish periodic from chaotic switching. Conversely, it divides the periodic non-switching phase into two physically different states, whose dynamics and site-resolved intensities are shown in Fig.~\ref{fig:modes} of Appendix~\ref{app:admd}. In the first, found mainly in the narrow strip near $\Delta\approx-3.5$, the two counter-propagating fields oscillate in exact synchrony, $I_1(t)=I_2(t)$ to machine precision, so the mirror symmetry is preserved. In the second, at $\Delta\approx5.5$--$7.5$ and strong drive, the symmetry is broken: the $+$ field dominates on the $A$ sites and the $-$ field on the $B$ sites, as expected from the cross-coupling of $a_n^{\pm}$ to $b_n^{\mp}$ in Eqs.~(\ref{eq:eom_a}) and (\ref{eq:eom_b}), and the oscillation is about eight times faster. The heuristic assigns both states to one class because a symmetric oscillation can never switch, whereas the library separates them without being told what to look for; the symmetry probe of the aDMD + GMM method separates them as well. Where the library disagrees with the reference on a large phase, the library can be the more accurate of the two: of the 26 points that the reference labels periodic but the library assigns to its bistable regime, 22 are bistable once the integration is extended to $t_f=240$. Against the $t_f=240$ heuristic the library reaches ${\rm ARI}=0.925$, and against a short-window heuristic ($t_f=60$, first third discarded) only $0.845$, which shows how strongly an unconverged reference can understate the agreement.

The aDMD + GMM method [Fig.~\ref{fig:diagrams}(d)] uses 9 of its 30 available mixture components and reaches ${\rm ARI}=0.947$, with homogeneity $0.899$ and completeness $0.839$. Other random initialisations and priors give 9 to 12 clusters at essentially the same ARI (Appendix~\ref{app:gmm}), so the number of clusters should not be read as a count of phases. The stable phase falls almost entirely into one cluster ($99\%$ of its points) and the bistable phase mostly into one ($94\%$), with a second, purely bistable cluster of 31 points that lie mostly along its edges. The symmetry probe places all 57 symmetric oscillations of the strip near $\Delta\approx-3.5$ in one cluster, which otherwise contains only 8 stable points, and 179 of the 183 symmetry-broken periodic non-switching points in another. Like the library, however, this method largely merges periodic and chaotic switching: one cluster contains 216 of the 218 chaotic switching points together with 92 of the 130 periodic switching points, and only 37 periodic switching points form a cluster of their own. The rare chaotic non-switching phase is not resolved as a separate cluster. The floor on the reconstruction error is essential: without it, the stable and bistable phases are each split roughly in half according to round-off-level differences in this error, and the agreement falls to ${\rm ARI}=0.53$ (Appendix~\ref{app:gmm}).

\rihead{Discussion}
The comparison shows that the four methods answer different questions rather than ranking along a single axis. The CNN gives the closest match to the reference, but only because it is trained on it: it reproduces a known classification quickly and with boundary-limited errors, yet it inherits any error in the training labels and cannot report a phase it has not been shown. The two label-free methods need no prior description of the phases and recover the main boundaries without labels, but their clusters must be interpreted afterwards, and their number is set by tolerances or by the statistical model rather than by the physics. They also arrive at similar partitions, but by different routes. Both separate the stable and bistable phases and the symmetric from the symmetry-broken oscillations, and both largely merge periodic and chaotic switching. The library compares bases between trajectories and fragments the oscillatory phases unless several discovery orders are combined, whereas aDMD + GMM summarises each trajectory by four numbers and returns a few large clusters, but its agreement relies on removing round-off from its reconstruction-error feature. Because the ARI is not linear in the fraction of misassigned points and counts pairs, so that large phases dominate it, the difference of 0.03 in ARI between the two label-free methods says less than the cluster-by-cluster comparison above.

For the treatment of a new lattice, these results suggest a workflow in which a label-free DMD pass is used first to locate the main boundaries and to flag regions with unexpected structure, a small number of points from each region is then inspected and labelled, and a supervised classifier is finally used to extend the labelled diagram to larger or finer parameter sweeps. This workflow is also affordable. On a reduced $20\times20$ grid and a single CPU core, the heuristic takes 2~s, aDMD + GMM 24~s, the aDMD library with three discovery orders 30~s, the CNN 54~s including simulation of its training set and training, and the autoencoder baseline 174~s (Appendix~\ref{app:cost}). The cost of the library grows faster than linearly with the number of grid points, because each trajectory is compared with every basis already in the library.

Two further lessons apply beyond this model. First, a classifier can only be judged against a reference that is itself converged. Lengthening the integration window from $t_f=60$ to $t_f=120$ changed $7\%$ of the heuristic labels, mostly because slowly decaying oscillations had been labelled periodic, and the label-free library had been penalised for classifying those trajectories correctly. Second, physically motivated features matter more than model complexity here. The ConvAE + IDEC baseline, which learns its own representation from the raw intensities and is even given the number of phases, reaches only ${\rm ARI}=0.375$ (Table~\ref{tab:comparison}), well below either DMD-based method. Feeding the four invariants of method (iv) into an autoencoder would add little, since they are already a four-dimensional representation that a GMM can cluster directly. A more promising direction is to learn representations from richer DMD outputs, such as the full eigenvalue spectrum or the library's reconstruction errors, or to use variational autoencoders and sequence models designed for time series; we leave these for future work. Both DMD-based methods also remain interpretable, since their features are frequencies, growth rates, and symmetry measures of the dynamics itself~\cite{wong2023,brunton2022}, whereas the internal representation of the CNN and of the autoencoder has no direct physical meaning.

\rihead{Conclusion}
We have compared a heuristic classifier, a supervised CNN, and two label-free DMD-based methods on the same driven, Kerr-nonlinear SSH lattice, using a converged heuristic diagram as the reference. The CNN reproduces the reference with $97.96\%$ held-out accuracy, with almost all errors on phase boundaries. Without any labels, the aDMD library and the aDMD + GMM method reach ${\rm ARI}=0.916$ and $0.947$ and recover the main phase boundaries, far better than an autoencoder baseline. The library also separates symmetric from symmetry-broken periodic oscillations, a physical distinction that the six-class heuristic misses, and in several regions where it disagrees with the reference a longer simulation shows it to be correct. Its main limitations are that it merges periodic and chaotic switching and that single discovery runs fragment oscillatory phases, which a consensus over discovery orders largely removes. Together, these results support using label-free DMD methods as a first survey of an unexplored nonlinear lattice, followed by targeted labelling and supervised classification. Looking forward, the same approach could be applied to two-dimensional topological lattices and other nonlinear photonic systems whose phase content is not known in advance~\cite{gong2020,dikopoltsev2021,wong2025npt}.

\rihead{Data availability}
The data that support the findings of this study are available from the corresponding author upon reasonable request.

\rihead{Code availability}
The code used for the simulations and the classification methods is available from the corresponding author upon reasonable request.

\rihead{Author contributions}
C.P.I. performed the simulations, implemented and analysed the classification methods, and wrote the manuscript. G.H.A. developed the driven Kerr SSH lattice model. S.W. developed the aDMD-based classification approach on which methods (iii) and (iv) are based and advised on its use. S.S.O. conceived and supervised the project. All authors discussed the results and contributed to the revision of the manuscript.

\rihead{Competing interests}
The authors declare no competing interests.

\clearpage
\onecolumngrid

\appendix

\setcounter{figure}{0}
\setcounter{table}{0}
\renewcommand{\thefigure}{S\arabic{figure}}
\renewcommand{\thetable}{S\arabic{table}}

\section{Numerical integration and convergence of the reference}
\label{app:conv}
All trajectories are integrated with a fixed-step fourth-order Runge-Kutta scheme with $\delta t=0.02$ up to $t_f=120$, in units of the inverse decay rate. Each run starts from $a_0^{+}=10^{-6}$, $a_0^{-}=0.99\times10^{-6}$, and all other amplitudes zero. The first half of each trajectory is discarded, and the remaining 3000 steps are sampled every third step, giving 1000 snapshots per trajectory. The parameter grid has $60\times60$ points, uniformly spaced over $(\Delta,I_s)\in[-10,10]\times[0,50]$, and all methods use these settings.

\begin{figure*}[h]
\centering
\includegraphics[width=\textwidth]{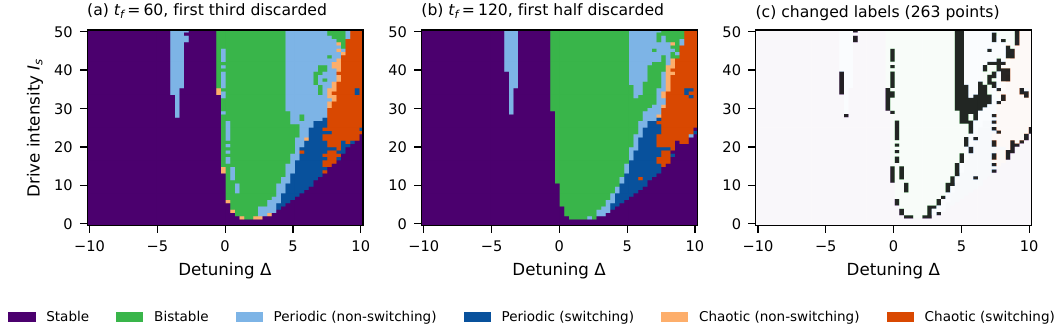}
\caption{Convergence of the heuristic reference. (a) Heuristic diagram from the short window ($t_f=60$, first third discarded). (b) Heuristic diagram from the window used in this work ($t_f=120$, first half discarded). (c) Grid points whose label differs between (a) and (b), shown in black over a faded copy of (b).}
\label{fig:conv}
\end{figure*}

The integration window was chosen by running the heuristic classifier (Appendix~\ref{app:heur}) with three windows: $t_f=60$ with the first third discarded, $t_f=120$ with the first half discarded, and $t_f=240$ with the first half discarded. The $t_f=120$ and $t_f=240$ diagrams agree on $98.4\%$ of the grid points (ARI $0.978$), so we regard $t_f=120$ as converged. The short-window $t_f=60$ diagram agrees with $t_f=120$ on only $92.7\%$ of the points (ARI $0.907$). As Fig.~\ref{fig:conv} shows, the 263 changed points lie almost entirely inside and along the edges of the bistable wedge: 182 of them are labelled periodic non-switching with $t_f=60$ but bistable with $t_f=120$. These trajectories are slowly decaying oscillations that have not yet reached their symmetry-broken steady state by $t=60$. A reference built from the shorter window would therefore penalise any method that correctly identifies them as bistable. Using one window for every method avoids comparing classifications made on different stages of the same trajectories.

\section{Heuristic classification rules}
\label{app:heur}
The heuristic classifier integrates a second trajectory alongside the first, started from slightly displaced initial amplitudes ($a_0^{+}=1.001\times10^{-6}$, $a_0^{-}=1.002\times10^{-6}$), and renormalises their separation to $D_0=10^{-6}$ after every step, following Benettin~\textit{et al.}~\cite{benettin1980}. The mean logarithmic growth of the separation per integration step, accumulated over the sampled window, is used as the estimate of the largest Lyapunov exponent; a trajectory is labelled chaotic if it exceeds $5\times10^{-3}$ per step, i.e.\ $0.25$ in units of the decay rate. For non-chaotic trajectories, the intensities $I_1$ and $I_2$ at the driven site are tested for constancy: an intensity counts as constant if its standard deviation over the window is below $10^{-6}$ or below $10^{-3}$ of its mean. A trajectory is labelled periodic if at least one intensity is not constant and the two intensities together show at least two local maxima that rise more than $10^{-5}$ above both neighbouring samples. Constant trajectories are labelled bistable if their mean intensities differ by more than $5\%$ of their average, and stable otherwise. Switching is detected from the sign of $I_1-I_2$, which is counted only when $|I_1-I_2|$ exceeds $10\%$ of the mean intensity; a periodic or chaotic trajectory is labelled switching if this sign changes at least twice within the window.

\section{aDMD library building}
\label{app:admd}
The library pipeline follows Refs.~\cite{wong2023,wong2026}. For each trajectory, DMD is computed from its last 800 snapshots (raw complex amplitudes of all 14 field components, without subtracting the temporal mean, which would remove the constant part that defines the stable and bistable states). The rank is chosen to retain $99\%$ of the singular-value sum, and each DMD mode $\psi$ is extended to $(\psi,\lambda\psi,\dots,\lambda^{N_w}\psi)$ with $N_w=25$ to form the augmented basis, which is orthonormalised. To test whether a basis describes another trajectory, five windows of $N_w+1$ consecutive snapshots are drawn at random from the last $30\%$ of that trajectory, and the median relative reconstruction error over the windows is used. In the bottom-up pass a new basis is added when this error exceeds $\epsilon_{\rm th}=0.02$ for every basis already in the library. In the top-down pass, two bases $Q_i$ and $Q_j$ are linked if their subspace overlap $\gamma_{ij}=\|Q_i^\dagger Q_j\|_F^2/\sqrt{r_ir_j}$ exceeds $\gamma_{\rm th}$, and linked bases are merged transitively. The threshold $\gamma_{\rm th}$ is taken as the smallest value in $\{0.90,0.93,0.95,0.96,0.97,0.98\}$ for which the largest merged regime covers less than $60\%$ of the grid; smaller values chain the stable and bistable regions together into a single regime. This criterion selected $\gamma_{\rm th}=0.98$, $0.97$, and $0.93$ for the three discovery orders used here, giving 187, 76, and 24 regimes, respectively.

The three partitions are combined by evidence accumulation~\cite{fred2005}. For every pair of grid points we count in how many of the three runs they share a regime; pairs that share a regime in at least two runs are linked, and the connected components of the resulting graph form the consensus regimes. The individual runs reach ${\rm ARI}=0.914$, $0.955$, and $0.891$ against the reference, and $0.910$--$0.951$ against one another; the consensus reaches $0.916$.

The run with the strictest threshold ($\gamma_{\rm th}=0.98$) shows how far the library can fragment. It assigns 116 grid points to regimes of their own, and $30\%$, $41\%$, and $35\%$ of the periodic non-switching, periodic switching, and chaotic switching points, respectively, fall into regimes of at most three points. For the stable and bistable phases this fraction is below $1\%$. Within the periodic phases, the regimes of this run are narrow bands of oscillation frequency: the dominant frequency of $I_1(t)$ varies with a standard deviation of $0.53$ across the periodic non-switching phase, but only $0.05$ on average within one of its regimes (regimes of at least five points). The bases of two periodic trajectories with different frequencies span different subspaces, so a strict merging threshold separates a continuous family of oscillations into many regimes. Chaotic trajectories have no low-rank DMD description at all, so the basis built from one chaotic trajectory rarely reconstructs another, and many of them start new regimes. Neither effect corresponds to a separate dynamical phase.

\begin{figure*}[t]
\centering
\includegraphics[width=\textwidth]{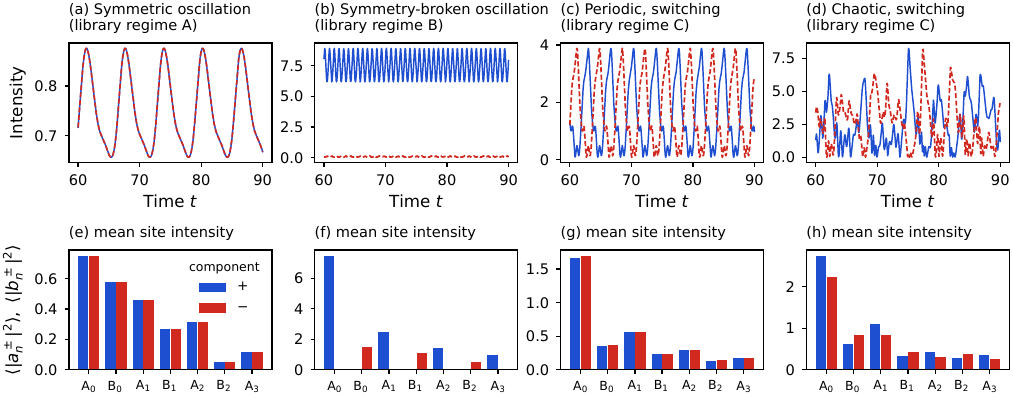}
\caption{Representative dynamics (top) and time-averaged site-resolved intensities $\langle|a_n^{\pm}|^2\rangle$ and $\langle|b_n^{\pm}|^2\rangle$ (bottom) for consensus regimes of the aDMD library. (a,e) Regime A, symmetric oscillation, $(\Delta,I_s)=(-3.56,41.53)$. (b,f) Regime B, symmetry-broken oscillation, $(5.93,43.22)$. (c,g) Regime C, periodic switching state, $(6.27,16.10)$. (d,h) Regime C, chaotic switching state, $(8.98,32.20)$. In the top row, $I_1$ is solid and $I_2$ dashed; in (a) the two coincide.}
\label{fig:modes}
\end{figure*}

The consensus keeps two divisions that do correspond to physically different states. Figure~\ref{fig:modes} shows representative trajectories and time-averaged site intensities for the two regimes into which the library divides the periodic non-switching phase, and for two members of its single switching regime. Regime A [Fig.~\ref{fig:modes}(a,e)], mainly in the strip near $\Delta\approx-3.5$, oscillates with $I_1(t)=I_2(t)$ to machine precision and has identical $+$ and $-$ intensity profiles along the chain. Regime B [Fig.~\ref{fig:modes}(b,f)], at $\Delta\approx5.5$--$7.5$, oscillates about eight times faster with strongly broken symmetry, the $+$ field being concentrated on the $A$ sites and the $-$ field on the $B$ sites. The heuristic gives both the label periodic non-switching, because a state that is never asymmetric cannot switch. Regime C [Fig.~\ref{fig:modes}(c,d,g,h)] contains both the periodic and the chaotic switching states, which the library does not separate.

\section{aDMD + GMM}
\label{app:gmm}
For each trajectory, the aDMD basis $Q$ is built exactly as for the library (Appendix~\ref{app:admd}): DMD of the last 800 snapshots, with the rank chosen to retain $99\%$ of the singular-value sum, extended to $(\psi,\lambda\psi,\dots,\lambda^{N_w}\psi)$ with $N_w=25$ and orthonormalised. The first invariant is $\log_{10}$ of the median relative reconstruction error $\sqrt{\|y\|^2-\|Q^\dagger y\|^2}/\|y\|$ over five windows $y$ of $N_w+1$ consecutive snapshots drawn at random from the last $30\%$ of the same trajectory. The second and third are $\log_{10}$ of the mean $|{\rm Im}\,\omega|$ and of the mean $|{\rm Re}\,\omega|$ of the continuous-time eigenvalues $\omega=\ln\lambda/\Delta t_{\rm s}$ of the same DMD, each weighted by the mode amplitudes that reconstruct the mean state. The fourth is $\log_{10}$ of the relative distance between the intensity vectors of the trajectory and of its counterpart started from the mirrored initial condition, each averaged over the second half of the sampled window. A constant of $10^{-6}$ is added before taking the logarithm of the second, third, and fourth invariants.

Because the reconstruction error is obtained from a difference of squared norms, it cannot resolve relative errors much below $\sqrt{\epsilon_{\rm mach}}\approx1.5\times10^{-8}$, where $\epsilon_{\rm mach}$ is the double-precision machine epsilon. For $98\%$ of the stable and bistable trajectories the error is below $10^{-6}$: it is exactly zero for $54\%$ of them and, with few exceptions, lies between $10^{-8}$ and $10^{-7}$ for the rest. These values are set by round-off rather than by the dynamics and lie more than four orders of magnitude below the library tolerance $\epsilon_{\rm th}=0.02$. Without a floor, the logarithm (with zero errors mapped to $10^{-12}$) turns this round-off into two well-separated groups, and the GMM splits both the stable and the bistable phase roughly in half, reaching only ${\rm ARI}=0.53$; a floor of $10^{-8}$, which lies below the round-off values, gives $0.61$. We therefore floor the error at $10^{-6}$. For floors between $10^{-7}$ and $10^{-3}$, the method returns 9 to 12 clusters with ${\rm ARI}=0.945$--$0.950$.

\begin{figure}[h]
\centering
\includegraphics[width=0.8\linewidth]{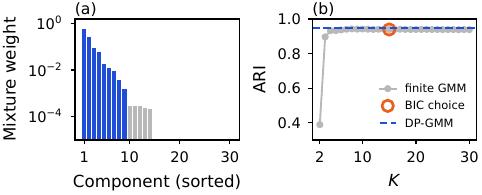}
\caption{aDMD + GMM model selection. (a) Mixture weights of the Dirichlet-process GMM, sorted; blue components are assigned at least one grid point and grey ones none; the remaining 16 components have weights below $10^{-5}$. (b) ARI against the reference. Dashed: the Dirichlet-process GMM used in this work (9 clusters). Grey: standard GMMs with a fixed number of components $K$, shown for comparison; the circle marks the $K=15$ that the BIC would select.}
\label{fig:bic}
\end{figure}

The invariants are standardised to zero mean and unit variance and clustered with a variational Dirichlet-process GMM~\cite{blei2006} with full covariance matrices, truncated at 30 components, with concentration prior $1/30$ and five random initialisations. The number of clusters is the number of components to which at least one grid point is assigned; Fig.~\ref{fig:bic}(a) shows the mixture weights. With truncation at 20 or 40 components, concentration priors between $10^{-3}$ and 1, or other random initialisations, the method returns 9 to 12 clusters with ${\rm ARI}=0.946$--$0.948$. For comparison, finite full-covariance GMMs with five random initialisations were fitted for $K=2$ to $30$ [Fig.~\ref{fig:bic}(b)]. Their ARI lies between $0.937$ and $0.943$ for every $K$ from 6 to 30, and the BIC selects $K=15$ with ${\rm ARI}=0.940$; a finite GMM selected by BIC therefore gives the same agreement with more clusters, seven of which are small groups of mostly boundary points.

\section{Supervised CNN}
\label{app:cnn}
The time-series branch receives the 14 field intensities, resampled to 256 points. Each channel is standardised to zero mean and unit variance; channels that are constant to within $10^{-6}$ of their mean are set to zero, because standardising them would amplify numerical round-off into apparently random signals. The branch has three one-dimensional convolutional layers (32, 64, and 128 channels, kernel sizes 7, 5, and 5, with batch normalisation and ReLU activation) followed by global average pooling. The spectral branch receives, for each channel, the logarithm of the windowed power spectrum averaged over 32 frequency bins, and passes it through two fully connected layers (128 and 64 units). The scalar branch receives the two mean intensities at the driven site, normalised by their average, and their normalised difference, and maps them to 16 units. The concatenated 208 features pass through a fully connected layer of 128 units with dropout to the six-class output. The network is trained for 100 epochs with AdamW (learning rate $3\times10^{-3}$, weight decay $10^{-4}$, batch size 64) and a cosine learning-rate schedule. Each training sample is weighted by the inverse frequency of its class, and label smoothing of $0.05$ is applied to the target distribution. The class weight must multiply the loss of each sample as a whole: in the built-in PyTorch loss, combining class weights with label smoothing also weights the smoothing term by class, which lets the rarest class (chaotic non-switching, 11 training points, against 1396 for the stable phase) dominate the loss of every sample.

\begin{figure}[h]
\centering
\includegraphics[width=0.6\linewidth]{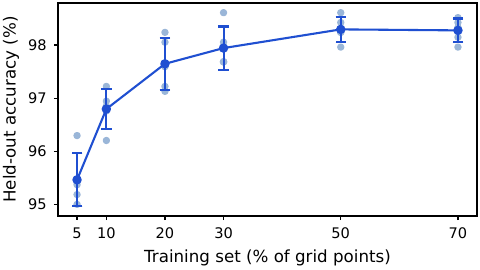}
\caption{Held-out accuracy of the supervised CNN versus training-set size. The test set is the same 1080 grid points for all networks. Light points show the five networks trained on independently drawn random subsets of each size; dark points and error bars show their mean $\pm$ one sample standard deviation.}
\label{fig:frac}
\end{figure}

To test how much labelled data the network needs, we kept the 1080 held-out points fixed and trained the network on random subsets of the remaining 2520 points, containing $5\%$, $10\%$, $20\%$, $30\%$, $50\%$, or $70\%$ of the grid (180 to 2520 points). For each size, five subsets were drawn with different random seeds, the first being the subset used in the main text, and each network was trained with the settings above. Figure~\ref{fig:frac} shows the resulting held-out accuracy. It saturates at $98.3\%$ for training sets of $50\%$ and $70\%$ of the grid and is $97.9\%$ for $30\%$, so the $70\%$ split of the main comparison is not necessary for this lattice. With $5\%$ of the grid labelled, the mean accuracy is still $95.5\%$. Across all 30 trained networks, $93\%$ of the misclassified points lie next to a phase boundary of the reference. The subset used in Fig.~\ref{fig:diagrams}(b) gives $97.96\%$, the lowest of the five values obtained with $70\%$ of the grid.

The overall accuracy is dominated by the large phases. The stable and bistable phases make up $84\%$ of the held-out points and are recognised with a recall of about $0.99$ for every training-set size, whereas the periodic and chaotic phases together make up only $16\%$ and are affected much more by the number of labels. For the subsets drawn with the first seed, the recall of the periodic non-switching phase falls from $0.96$ with $70\%$ of the grid labelled to $0.69$ with $5\%$, that of the chaotic non-switching phase from $0.67$ to $0.17$ (only 6 held-out points), and the balanced accuracy, i.e.\ the mean recall over the six phases, from $0.92$ to $0.77$. With $5\%$ of the grid labelled, the chaotic non-switching phase is represented on average by fewer than one training point.

\section{ConvAE + IDEC baseline}
\label{app:ae}
The baseline simulates each grid point from three initial conditions (the default one with its overall amplitude rescaled by a random factor $1+0.1\xi$, $\xi\sim\mathcal N(0,1)$), resamples the 14 intensities to 256 points, and standardises each sample with a single mean and standard deviation over all channels, which preserves the intensity difference between the two field components. A one-dimensional convolutional autoencoder with a 32-dimensional latent space is trained on all 10800 sequences for 100 epochs, after which the improved deep embedded clustering (IDEC) loss~\cite{guo2017} with six clusters is added to the reconstruction loss for 40 further epochs. Each grid point receives the majority label of its three sequences. The baseline is therefore told the number of reference phases, which the DMD-based methods are not.

\section{Computational cost}
\label{app:cost}
For an indicative comparison, each pipeline was run with the settings above on a reduced $20\times20$ grid (400 points) on a single CPU core, with all numerical libraries restricted to one thread. The CPU times were 2~s for the heuristic classifier, 24~s for aDMD + GMM (including the second, mirrored simulation of every point and the Dirichlet-process GMM fit), 30~s for the aDMD library with three discovery orders, 54~s for the CNN (simulation and feature extraction of the 280 training points, 100 training epochs, and prediction of the other 120 points), and 174~s for the ConvAE + IDEC baseline (three simulations per point). These times exclude one-off compilation. Absolute values depend on hardware and implementation, and the costs do not all scale in the same way with grid size. The library's discovery and classification steps compare each trajectory with every basis or regime, so their cost grows faster than linearly with the number of points, whereas the heuristic and aDMD + GMM costs grow linearly and the CNN cost is dominated by training.

\end{document}